\documentclass[prl,twocolumn]{revtex4}

\usepackage{bm}

\begin{document}

\title{A Comment on Liu and Yau's positive quasi-local mass}

\author{N.\'{O} Murchadha}
\email{niall@ucc.ie}
\affiliation{Physics Department University College Cork, Ireland}

\author{L.B. Szabados}
\email{lbszab@rmki.kfki.hu}
\affiliation{Research Institute for Particle and Nuclear 
          Physics, H--1525 Budapest 114, P. O. Box 49, Hungary}

\author{K.P. Tod}
\email{tod@maths.ox.ac.uk}
\affiliation{Mathematical Institute, Oxford OX1 3LB, England}

\begin{abstract}
With the aid of a simple family of examples, we show that the 
quasi-local mass defined by Kijowski \cite{K} and Liu and Yau 
\cite{LY} and shown by Liu and Yau to be positive, 
may be strictly positive for space-like, topologically spherical 2-surfaces 
in flat space-time.

\end{abstract}

\maketitle

In a recent letter, \cite{LY},  Liu and Yau propose a 
definition of quasi-local mass for any space-like, topological 2-sphere 
with positive intrinsic (or Gauss) curvature. 
In fact, their expression is precisely the `quasi-local free 
energy' of Kijowski \cite{K}. 
However, Liu and Yau are able to 
prove positivity of this mass by exploiting a result of Shi and Tam, 
\cite{ST}, which proves positivity of the ADM mass on space-like 
3-surfaces with an inner boundary. This is an impressive result. 

As Liu and Yau note, there have been many attempts to define a 
quasi-local mass or energy associated to a space-like 2-surface and 
there are certain natural questions to ask of any new definition. 
One possible set of such questions was published by Christodoulou 
and Yau, \cite{CY}, as properties which a definition \emph{must} have. 
There is no general consensus as to the necessity of all the 
properties demanded by Christodoulou and Yau, but the first of them, 
that the definition should give zero for any 2-surface in flat 
space-time, has wide support. It is therefore worth noting, and it 
is the purpose of this Comment to note, that the definition of 
Kijowski, Liu and Yau does not have this property: there are space-like, 
topological 2-spheres in flat space for which their mass is strictly 
positive. In fact there is a simple class of examples for which the 
whole construction is easy to see and we shall present these below.

The definition of mass given by Kijowski, Liu and Yau belongs to the 
class of definitions initiated by Brown and York \cite{BY} for a 
2-surface $\Sigma$ 
of the kind considered. A definition in this class considers a quantity 
$Q$ obtained by integrating over $\Sigma$ an expression constructed from 
the \emph{data} of $\Sigma$, by which we mean the first and second 
fundamental forms and the connection on the normal bundle. To complete 
the definition, one then subtracts a quantity $Q_0$  from $Q$ which is 
intended to be the value `in flat space-time' of $Q$. Difficulties 
arise with saying precisely what the subtraction should be, as a 
2-surface in curved space-time cannot in general be embedded in flat 
space-time with the same data. 

Kijowski, Liu and Yau restrict consideration to 2-surfaces $\Sigma$ 
with positive intrinsic curvature (and subject to a second condition, 
given below). By Weyl's embedding theorem, such a $\Sigma$ can be 
isometrically embedded in flat (Euclidean) 3-space $E^3$ with 
positive-definite extrinsic curvature tensor. 
The embedding is unique up to isometry of $E^3$. The flat $E^3$ can be 
thought of as a constant-time hyperplane in flat (Minkowski) space-time, 
and it is then this embedding which is used to define the subtraction 
mentioned above. Specifically, the Kijowski-Liu-Yau definition of 
quasi-local energy (equation (105) of \cite{K} and (3) of \cite{LY}) is 
\begin{equation}
E_{KLY} = \frac{1}{8\pi G}\int_\Sigma (k_0-k)
\label{1}
\end{equation}
where $G$ is Newton's gravitational constant, $k_0$ is the trace of the 
extrinsic curvature of $\Sigma$ as embedded in $E^3$ and $k$ is given in 
terms of the Newman-Penrose spin-coefficients $\rho$ and $\mu$, (see 
\cite{NP} or \cite{PR}), as
\begin{equation}
k=\sqrt{8\rho\mu}.
\label{2}
\end{equation}
Here, as is standard, $\rho$ is minus half the expansion of the 
outgoing future null normal to $\Sigma$, and $\mu$ is half the 
expansion of the ingoing future null normal, suitably normalised. 
Kijowski 
and Liu and Yau require the product $\rho\mu$ to be positive 
for their definition, which is equivalent to the condition that the 
mean-curvature vector of $\Sigma$ be time-like. This is the second 
condition on $\Sigma$ whose necessity was noted above.

Clearly, equation (\ref{1}) will give $E_{KLY}=0$ for any $\Sigma$ 
which lies in a space-like hyperplane in flat space-time. However, for 
a $\Sigma$ in flat space-time, with positive intrinsic curvature and 
time-like mean curvature vector but which does \emph{not} lie in a 
space-like hyperplane, there is no reason to think that $E_{KLY}$ 
should vanish, and in fact it need not. Our purpose now is to describe 
a simple class of $\Sigma$ with these properties and non-zero $E_{KLY}$.

The idea is to consider a 2-surface which is a cross-section of the 
light-cone of the origin in flat space-time. We introduce spherical 
polar coordinates $(t,r,\theta,\phi)$ based at the origin so that the 
Minkowski metric is 
$$
ds^2=dt^2-dr^2-r^2(d\theta^2+\sin^2\theta\, d\phi^2).
$$
The future light-cone $N$ of the origin is given by $t=r>0$ and we may 
define a topological 2-sphere $\Sigma$ lying on $N$ by an equation of 
the form 
\begin{equation}
t=r=F(\theta,\phi)
\label{F}
\end{equation}
for a positive, smooth function $F$. With a natural choice of normals 
to $\Sigma$, the relevant spin-coefficients are given by 
\begin{eqnarray}
\rho & = & \frac{-1}{\sqrt{2}F}\\
\mu & = & \frac{-1}{\sqrt{2}F}(1-\Delta\log F)
\end{eqnarray}
where $\Delta$ is the Laplacian on the unit sphere. The spin-coefficient 
$\sigma$ is zero (as $\Sigma$ lies on a light-cone and these are shear-free 
in flat space-time) so that the intrinsic curvature $K$ of $\Sigma$, 
calculated for example from the formula (4.14.20) in \cite{PR}, is 
given by 
\begin{equation}
K=2\rho\mu.
\label{K}
\end{equation}
(Note the relation of $K$ to $k$ in equation (\ref{2}): $k=2\sqrt{K}$). 

Thus $\Sigma$ has positive intrinsic curvature and time-like mean 
curvature vector iff $\Delta\log F<1$, which is easy to arrange. 
Suppose this holds, then following the prescription of Kijowski, Liu 
and Yau, we embed $\Sigma$ into $E^3$, where it will define a convex 
topological 2-sphere $\Sigma_0$ with mean curvature say $H$ and 
intrinsic curvature 
$K$ as above. (Conversely, we can obtain a $\Sigma$ subject to the 
conditions considered by starting from such a $\Sigma_0$.) We calculate 
$E_{KLY}$ using equations (\ref{1}), (\ref{2}) and (\ref{K}) above to 
find the expression
\begin{equation}
E_{KLY}=\frac{1}{8\pi G}\int_\Sigma (2H-2\sqrt{K}).
\label{ans}
\end{equation}
At this stage, we are considering the expression (\ref{ans}) for an 
arbitrary convex topological 
2-sphere $\Sigma_0$ in $E^3$, and clearly it need not 
vanish. If we introduce the usual principal curvatures $\lambda_1$ and 
$\lambda_2$, both of which are positive for a convex surface, then 
(\ref{ans}) becomes 
\[E_{KLY}=\frac{1}{8\pi G}\int_\Sigma(\sqrt\lambda_1-\sqrt\lambda_2)^2.\]
This vanishes for a round 2-sphere (as expected, since in this case the 
cross-section of $N$ in fact lies in a hyperplane) but is otherwise 
positive: the Kijowski-Liu-Yau definition can be strictly positive for 
space-like, topological 2-spheres in flat space-time.

Kijowski has another definition for the quasi-local energy (equation 
(97) of \cite{K}), which is also positive for the 2-surfaces presented 
here. The Brown-York energy with the flat space reference \cite{BY} and 
the two light cone references considered in \cite{BLY} and in \cite{L} 
are also non-zero in general. 

This work was initiated at the Erwin Schr\"odinger Institut, Vienna 
and we gratefully acknowledge the hospitality of the Institut. 
LBSz was partially 
supported by the Hungarian Scientific Reseach Fund 
grants OTKA T030374 and T042531.

\end{document}